\documentstyle[twocolumn,aps,epsfig,floats]{revtex}
\begin{document}

\twocolumn[\hsize\textwidth\columnwidth\hsize\csname %
@twocolumnfalse\endcsname

\title{Theory of NMR as a local probe for the electronic
structure in the mixed state of the high-$T_c$ cuprates}
\author{Dirk K. Morr $^{1}$ and Rachel Wortis $^{2}$}
\address{ $^{1}$ University of Illinois at Urbana-Champaign,  
Loomis Laboratory of Physics, 1110 W. Green St., Urbana, IL 61801\\
$^{2}$ McMaster University, Dept. of Physics and Astronomy, 
1280 Main St. W., Hamilton, ON L8S-4M1}
\date{\today}
\draft
\maketitle
\begin{abstract}
We argue that nuclear magnetic resonance experiments are a
site-sensitive probe for the electronic spectrum in the mixed state of 
the high-$T_c$ cuprates. Within a spin-fermion model, we show that 
the Doppler-shifted electronic spectrum
arising from the circulating supercurrent changes the low-frequency
behavior of the imaginary part of the spin-susceptibility. For a hexagonal 
vortex 
lattice, we predict that these changes lead to {\it (a)} a unique dependence of 
the $^{63}$Cu spin lattice relaxation rate, $1/T_1$, on resonance frequency, 
and  
{\it (b)} a temperature dependence of $T_1$ which varies with frequency. We 
propose a nuclear quadrupole experiment to study the effects of a uniform 
supercurrent on the electronic structure and predict that $T_1$ varies with the 
direction of the supercurrent.

\end{abstract}


]

The form of the fermionic excitation spectrum in and around a vortex
in the mixed state of high-$T_c$ cuprates has been the subject of
intense research in the last few years \cite{Vol93,Wan95}. 
It has been known for a long time that the supercurrent circulating a
vortex gives rise to a Doppler-shift in the fermionic excitation
spectrum \cite{Tink80}. For a d-wave order parameter, Volovik pointed out 
\cite{Vol93}, this shift leads to
the scaling relation $D(\epsilon_F) \sim \sqrt{H}$, where
$D(\epsilon_F)$ is the density of states (DOS) at the Fermi energy,
and $H$ is the applied field. The main contribution to the DOS comes 
from regions close to the nodes, but in real space far from the 
vortex cores. Specific heat (SH) measurements by Moler {\it et
al.}~\cite{Mol94} on YBa$_2$Cu$_3$O$_{6.95}$ confirmed this scaling
behavior, however, these experiments only provide information on the sample 
averaged DOS.

In this communication we argue that the spin-lattice relaxation time,
$T_1$, measured in nuclear magnetic resonance (NMR) experiments, 
is a more direct probe since it can provide site-specific information on
the electronic structure in the vicinity of a vortex. 
Our conclusions are based on two important results. First, it was recently
pointed out by Curro and Slichter (CS) that the dependence of resonance 
frequency on the local field allows one to measure $T_1$ as a function of 
the nucleus position in a vortex lattice \cite{Cur98}. In addition,
$T_1$-measurements are a purely electronic probe, in contrast to SH 
experiments, which are dominated by phononic contributions.
Second, within a spin-fermion model \cite{Morr98}, we find that the 
Doppler-shifted electronic spectrum significantly changes the low-frequency
behavior of the imaginary part of the spin susceptibility,
$\chi''({\bf q}, \omega)$. Since $\chi''({\bf q}, \omega)$ changes with the 
spatially varying supercurrent, $1/T_1$  which is a weighted momentum
average of $\chi''$ in the zero frequency limit, acquires a frequency, 
i.e., spatial dependence.

Experimentally, CS found for the $^{17}$O resonance in
YBa$_2$Cu$_3$O$_7$ that $1/T_1$ increases with decreasing
distance from the vortex core \cite{Cur98}, i.e., with increasing
frequency. Preliminary results of Milling {\it et al.}~\cite{Mil99}
show a qualitatively similar behavior for the $^{63}$Cu relaxation
rate. We demonstrate that the experimentally observed 
frequency dependence of $T_1$ can be qualitatively explained by the spatial 
variation of the supercurrent and the resulting Doppler-shift in the electronic 
spectrum and predict a temperature dependence of the relaxation 
rate which changes uniquely with frequency. Both results provide insight into
the {\it local} structure of electronic and magnetic excitations in the mixed 
state, in contrast to the sample averaged information 
provided by SH experiments.

In what follows we consider for simplicity a single-layer system; an 
extension of our results to the case of a double-layer system, i.e., 
YBa$_2$Cu$_3$O$_{6+x}$, is straightforward. The $^{63}$Cu spin-lattice
relaxation rate, $1/T_1$, for an applied field parallel to the
$c-$axis is given by  
\begin{equation}
{1 \over T_1 T} = { k_B \over 2 \hbar} (\hbar^2 \gamma_n \gamma_e)^2 
{ 1 \over N} \sum_{\bf q} F_c({\bf q}) \lim_{\omega \rightarrow 0} 
{ \chi''({\bf q}, \omega) \over \omega}  \ ,
\label{T1T} 
\end{equation}
where 
\begin{equation}
F_c(q) =\Big[A_{ab}+2B \Big( cos(q_x)+cos(q_y) \Big) \Big]^2 \ ,
\label{form}
\end{equation}
and $A_{ab}$ and $B$ are the on-site and transferred hyperfine
coupling constants, respectively \cite{Zha96,Morr99}. To describe the effects 
of 
a
supercurrent on $\chi''$ we use the spin-fermion model \cite{Morr98}
in which the damping of the spin mode is determined by
its coupling to the planar quasi-particles. Within this model, the
spin propagator, $\chi$,  is given by  
\begin{equation}
\chi^{-1} = \chi_0^{-1} -  \Pi   \ ,
\label{Dyson}
\end{equation}
where $\chi_0$ is the bare propagator, and $\Pi$ is the bosonic
self-energy given by the irreducible particle-hole bubble. It was
earlier suggested \cite{MMP} that Re$\, \chi^{-1}$ should possess a
hydrodynamic form such that for $\omega \rightarrow 0$ 
\begin{equation}
{\rm Re}\, \chi^{-1} = \chi_0^{-1} - {\rm Re}\, \Pi 
= { \xi_{AF}^{-2} +  ({\bf q}-{\bf Q})^2 \over \alpha }  \ ,
\label{chifull}
\end{equation}
where $\xi_{AF}$ is the magnetic correlation length, $\alpha$ is 
a temperature independent overall constant, and ${\bf Q}=(\pi,\pi)$  is the
position of the magnetic peak in momentum space which we assume to be
commensurate \cite{com1}. 
The description of Re$\, \chi^{-1}$ in Eq.(\ref{chifull}) as  a
relaxational, non-dispersing spin mode at low frequencies is in
agreement with the analysis of NMR and inelastic neutron
scattering experiments on several La$_{2-x}$Sr$_x$CuO$_4$ and
YBa$_2$Cu$_3$O$_{6+x}$  compounds \cite{Morr99,MMP,Mas96}. 

We are thus left with the calculation of the imaginary part of $\Pi$
which describes the spin damping brought about by the decay of a spin
excitation into a particle-hole pair. In the superconducting state we
find to lowest order in the spin-fermion coupling $g$ 
\begin{eqnarray}
-\Pi({\bf q}, i \omega_n) &=& g^2 \, T \sum_{{\bf k},m} \ 
\Big\{ G({\bf k}, i\Omega_m) G({\bf k+q}, i\Omega_m+i\omega_n) 
\nonumber \\
& & + F({\bf k}, i\Omega_m) 
F({\bf k+q}, i\Omega_m+i\omega_n) \Big\} \ , 
\label{Pi}
\end{eqnarray}
where $G$ and $F$ are the normal and anomalous Green's functions.
Assuming that the supercurrent momentum, ${\bf p}_s$, possesses
only a weak spatial dependence, we find in semiclassical approximation
\cite{Tink80} 
\begin{eqnarray}
G({\bf k}, i\Omega_m) &=& { v^2_{\bf k} \over i\Omega_m - E_{\bf k} } +
{ u^2_{\bf k} \over i\Omega_m + E_{\bf - k} } \nonumber \ ; \\
F({\bf k}, i\Omega_m) &=&  -u_{\bf k} v_{\bf k} 
\left\{ {1 \over i\Omega_m - E_{\bf k} } 
        - { 1 \over i\Omega_m + E_{\bf - k} } \right] \ .
\label{GF}
\end{eqnarray}
Here, $E_{\bf k}$ is the dispersion of the fermionic
quasi-particles (Bogoliubons), which up to linear order in $p_s$ is
given by 
\cite{Tink80}
\begin{equation}
E_{\bf k} = \sqrt{  \epsilon_{\bf k}^2 + |\Delta_{\bf k}|^2} + 
{\bf v}_F({\bf k})\cdot{\bf p}_s \ ,
\label{qp_disp}
\end{equation}
where $\epsilon_{\bf k}$ is the electronic normal state dispersion and 
${\bf v}_F({\bf k})=\partial \epsilon_{\bf k} /\partial {\bf k}$.
Any changes in the  $d$-wave gap  
$\Delta_{\bf k}=\Delta_0 \   (\cos(k_x) - \cos(k_y))/2$ and the BCS
coherence factors $u_{\bf k}, v_{\bf k}$ due to a supercurrent  appear
only to order $p_s^2$ and can thus be neglected. 

Since the dominant contribution to Im$\, \Pi$ for $\omega \rightarrow
0$ comes from quasi-particle excitations in the vicinity of the nodes,
we expand $E_{\bf k}$ around the nodes to linear order in
momentum and perform the momentum and frequency integrations in Eq.(\ref{Pi}) 
analytically. Combining the results with Eqs.(\ref{T1T}),
(\ref{Dyson}) and  (\ref{chifull}) we find that the dependence of
$1/T_1$ on temperature $T$ and on ${\bf p}_s$ is determined by the set
$\{D_m/T\}$, where $D_m={\bf v}^{(m)}_F \cdot {\bf p}_s$ and ${\bf
v}^{(m)}_F$ is the Fermi velocity at the node in the $m$'th quadrant
of the Brillouin zone.   
The full expression for $1/T_1$ for arbitrary $\{D_m/T\}$ is too
cumbersome to be presented here, however,  
in the low temperature limit where $|D_m/T| \gg 1$ for $m=1..4$, it
simplifies and we obtain up to order $(T/D_m)^2$  
\begin{eqnarray}
{1 \over T_1 T} &=& 
{ {\cal C} \over N} \left({ 1  \over v_F v_\Delta }\right)^2
\sum_{n,m}  \Bigg\{ {\cal F}({\bf q}_{n,m}) \left(D_m \, D_n+{ \pi^2
T^2\over 3}  
\right) \nonumber \\
& & \hspace{-0.5cm} + {\cal F}({\bf q}_{n,m+2}) \left(D_m \, D_n-{
\pi^2 T^2\over 3} \right) \Bigg\} +O(e^{D_m/T}) 
\label{T1Tf1}
\end{eqnarray}
where 
the sum runs only over those nodes with
$D_m,D_n<0$,
${\cal C}= (\alpha g)^2 k_B  (\hbar^2 \gamma_n \gamma_e)^2/(2
\hbar)$, $v_\Delta=|\partial \Delta_{\bf k} /\partial {\bf k}|$ at the
nodes, and 
\begin{equation}
{\cal F}({\bf q}_{n,m})={F_c({\bf q}_{n,m})  \over (\xi_{AF}^{-2} + ({\bf
q}_{n,m}-{\bf Q})^2)^2} \ . 
\label{calF}
\end{equation} 
Here, ${\bf q}_{n,m}$ is the momentum connecting nodes $n$ and $m$. 
The opposite signs of the  $T^2$-terms on the r.h.s.~of
Eq.(\ref{T1Tf1}) arises from the combination of Fermi functions
involved in different relaxation processes. While the first term
describes a process in which a Bogoliubon is simply scattered at the
spin-fermion vertex, the second one  involves processes in which two
Bogoliubons are simultaneously created or destroyed  at the vertex.
Since the dominant contribution to the relaxation rate in the low
temperature regime comes from the second term in Eq.(\ref{T1Tf1}), we
find the unexpected result that  in this limit $1/T_1T$ actually 
decreases with increasing temperature. 
Note, that for $T \rightarrow 0$ it follows from
Eq.(\ref{T1Tf1}) that $1/T_1T \sim p_s^2$. 
 
In the high-temperature limit ($|D_m/T| \ll 1$ for all $m$), we obtain
to leading order in $(D_m/T)^2$ 
\begin{equation}
{1 \over T_1 T} = 
{ \pi {\cal C}  \over 6N} \left( {T \over 
v_F  v_\Delta } \right)^2 \sum_{n,m} \Big[ {\cal F}({\bf q}_{n,m})   + 
 {\cal F}({\bf q}_{n,m+2}) \Big]
\label{T1Tf2}
\end{equation}
where the sum runs over all nodes. For $T \gg |D_m|$ we thus recover
as expected the temperature dependence of the relaxation rate in the 
absence of a supercurrent. 

We now turn to NMR experiments in the mixed state of the high-$T_c$
cuprates which possess a hexagonal vortex lattice.  We consider for 
definiteness YBa$_2$Cu$_3$O$_7$, where $v_F \approx 0.4$ eV,  
$v_\Delta \approx 20$ meV, $A_{ab}/B = 0.7$, and $\xi_{AF} \approx 2$ is 
temperature independent in the superconducting state \cite{Morr99}. 
Each nucleus in the sample is
characterized by a resonance frequency $\Delta\nu({\bf r})= \gamma_n
\hbar H_z({\bf r})$ and a supercurrent momentum ${\bf p}_s({\bf r})$.
Here $H_z({\bf r})$ is the local magnetic field and $\gamma_n$ is
the nuclear gyromagnetic ratio. 
Since the electronic Zeeman-splitting for typical applied
fields is of the order $10^{-1}$ meV, and thus smaller than the
Doppler-shift for most of the nuclei, it can be neglected. 

Our scenario for the calculation of $T_1$ is only applicable to a 
given nucleus, if in its vicinity $\Delta({\bf r})$ is uniform and
${\bf p}_s({\bf r})$ varies sufficiently slowly. 
Since both assumptions are violated in the
vicinity of a vortex core we exclude in what follows all nuclei within
a radius $R=2\xi_{ab} \approx 6a_0$ of the vortex core. 
Here, $\xi_{ab}$ and $a_0$ are the superconducting in-plane coherence
length and the crystal lattice constant, respectively, and  
$H_z({\bf r})$ is given by \cite{Tink80} 
\begin{equation}
H_z({\bf r}) =  {2 \Phi_0 \over \sqrt{3} \lambda^2} \sum_{\bf q} {
exp(i{\bf q r}) \over 1 + \lambda^2 q^2 } 
\end{equation}
where $\lambda$ is the magnetic penetration depth, $\Phi_0$ is the flux 
quantum, and ${\bf q}$ runs
over the reciprocal lattice of the 2D vortex array. Since non-local
as well as non-linear effects can effectively be accounted for by a
renormalization of $\lambda$ \cite{Amin98} we neglect them in the
following.  We can then calculate the supercurrent momentum via
$\nabla \times {\bf H}({\bf r}) \sim {\bf p}_s({\bf r})$, and thus
obtain a spatial relation between $H_z({\bf r})$ and ${\bf p}_s({\bf r})$.  
\begin{figure} [t]
\begin{center}
\leavevmode
\epsfxsize=7.5cm
\epsffile{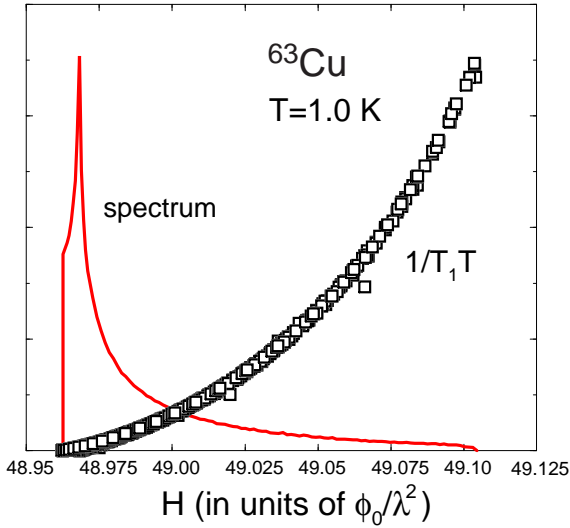}
\end{center}
\caption{The $^{63}$Cu spectrum (solid line) and $1/T_1T$ (open
squares) as a function of the local field for T=1 K. Both curves are
truncated at a distance $2\xi_{ab}$ from the vortex core.} 
\label{T1T_latt_spec}
\end{figure}
In Fig.~\ref{T1T_latt_spec} we plot the resulting $^{63}$Cu spectrum
(solid line) and $1/T_1T$ (open squares) as a function of the local field 
$H_z$ at $T=1$ K.   
Nuclei at the highest frequencies are located  at a distance $2\xi_{ab}$
from the center of the vortex, nuclei at the lowest frequencies are in
the center of a vortex triangle,  those at the maximum of the spectrum
are at the midpoint between two vortices.  For $T=1$ K, the relaxation
rate for practically all nuclei is determined by  
the low-temperature expression, Eq.(\ref{T1Tf1}). In this case,
$1/T_1T \sim p_s^2$ and hence reflects the frequency dependence of
$p_s$ which increases with decreasing distance from the vortex
core.  
The Doppler-shifted fermionic spectrum thus gives rise to a unique
dependence of $1/T_1T$ on frequency within the vortex-lattice
spectrum of $^{63}$Cu.
This result is in qualitative agreement  with preliminary
measurements by Milling {\it et al.}~\cite{Mil99} who indeed find
that $1/T_1T$ increases with increasing local field. 

\begin{figure} [t]
\begin{center}
\leavevmode
\epsfxsize=7.5cm
\epsffile{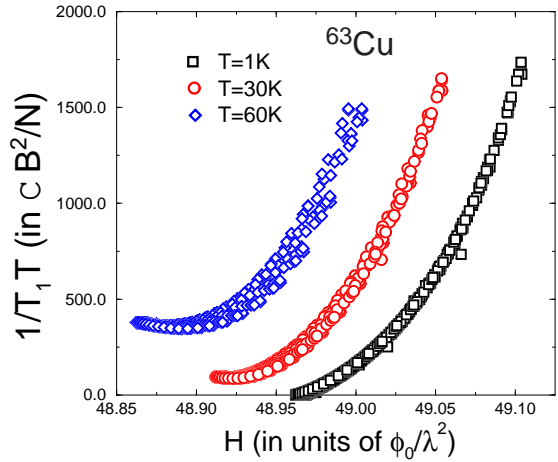}
\end{center}
\caption{$(1/T_1T)$ as a function of the local field for three 
different temperatures. The 30 K and 60 K curves are horizontally
offset for clarity.} 
\label{T1T_latt_T}
\end{figure}

Another unique signature of Doppler-shifted excitations appears in the
temperature dependence of the relaxation rate. 
In Fig.~\ref{T1T_latt_T} we present $1/T_1 T$ as a function of field
for three different temperatures (the 30 K and 60 K curves are 
horizontally offset for clarity).  
For nuclei at the low-field end of the curve, $p_s$ is small, $|D_m/T| \ll 1$ 
already at low T, and hence $(T_1T)^{-1}\sim T^2$ which increases with 
increasing temperature.  
On the other hand, for nuclei closer to  the vortex core, and
particularly  for those  at the high-field end of the curve, 
$p_s$ is large and  
$|D_m/T| \gg 1$ even for $T=60$ K (which is still below the lattice
melting temperature at $T_m=70$ K). The temperature dependence of
$1/T_1T$ is therefore described by Eq.(\ref{T1Tf1}), and thus
decreases with increasing temperature, as discussed earlier. 
Between these two extrema there exists a cross-over region, 
characterized by a minimum in the relaxation rate. This non-monotonic
behavior of $1/T_1T$ arises from two competing contributions: while
quasi-particle excitations involving nodes with $D_m>0$ become
exponentially suppressed  with increasing $H_z$, the contributions
from the remaining (unsuppressed) excitations scale as $ \sim p_s^2$
and thus increase. The cross-over occurs approximately at those
nuclei for which $max{|D_m|} \sim O(T)$; the minimum thus
shifts towards higher fields, i.e., larger values of $|D_m|$, with increasing 
temperature, as shown in Fig.~\ref{T1T_latt_T}. 

While the absolute scale of our results in Figs.~\ref{T1T_latt_spec}
and \ref{T1T_latt_T} depends on a number of parameters, it is practically 
independent of $\xi_{AF}$ for $\xi_{AF} >3$. In contrast, the predicted
temperature and frequency dependence of $T_1$ depends only on 
$\{D_m/T\}$ and thus presents a generic feature of a Doppler-shifted electronic 
spectrum due to the distribution of supercurrents
in a hexagonal vortex lattice.  

An alternative experiment which could observe the effects of a
Doppler-shifted fermionic dispersion is a nuclear quadrupole resonance (NQR) 
measurement in which a uniform supercurrent is applied to the sample. Since in 
the low-temperature limit, Eq.(\ref{T1Tf1}), 
$1/T_1T$ depends on $D_m={\bf v}^{(m)}_F \cdot {\bf p}_s$, with ${\bf
v}^{(m)}_F$ being fixed by the underlying lattice, one would expect
that the relaxation rate varies with the angle between the supercurrent 
momentum, ${\bf p}_s$, and the crystal axes.  
\begin{figure} [t]
\begin{center}
\leavevmode
\epsfxsize=7.5cm
\epsffile{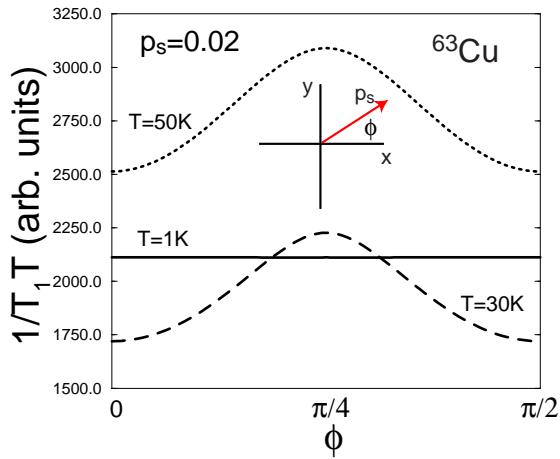}
\end{center}
\caption{$1/T_1T$ for $p_s=0.02$ and three different
temperatures as a function of the angle $\phi$ between ${\bf p}_s$ and
the crystal $x-$axis (see inset). } 
\label{T1T_phi}
\end{figure}
In Fig.~\ref{T1T_phi}  we plot $1/T_1T$, for $|{\bf p}_s|=0.02$ as a
function of the angle $\phi$ between ${\bf p}_s$ and the crystal
$x-$axis (see inset).  
For $\phi=0$, i.e., for ${\bf p}_s$ parallel to the $x-$axis, we find
$D_{2,3}=-v_F \, q_s/\sqrt{2}$ while  $D_{1,4}=+v_F \, q_s/\sqrt{2}$.
$1/T_1T$ thus decreases with T for $ |D_m/T| \gg 1$, and increases as
$\sim T^2$ for $|D_m/T| \ll 1$, in agreement with the results
in Fig.~\ref{T1T_phi}. 
On the other hand, $\phi=\pi/4$, i.e., ${\bf p}_s$ along the lattice
diagonal, is a special case since $D_{1,3}=\pm v_F \, q_s$ while
$D_{2,4}=0$.  
While the contribution to $1/T_1T$ from quasi-particle excitations
involving only nodes 2 and 4  scales as $T^2$ for all temperatures,
the contributions involving only nodes 1 and 3 decreases with
increasing T. As a result, $1/T_1T$  increases for all temperatures,
though only weakly at low T. For $T \gg v_F \, q_s $, the 
relaxation rate for all $\phi$ scales as $T^2$ and thus becomes
angle-independent.  
Note, that the strong angular dependence of the relaxation rate for
intermediate  
temperatures explains the spread in values of
$1/T_1T$ at a given local field which we found in
Fig.~\ref{T1T_latt_T}, particularly in the T=60K curve.

Spin-diffusion and vortex vibrations \cite{Dem94} can potentially
contribute to the spin-lattice relaxation and thus smear out the
effects described above. However, spin-diffusion is strongly
suppressed by the inhomogeneity of the magnetic field in a vortex
lattice \cite{Wor98}.
Moreover, calculations on the effect of vortex vibrations in 
YBa$_2$Cu$_3$O$_{6+x}$ \cite{Wor98}
as well as an experimental comparison of $^{17}$O and $^{63}$Cu
relaxation rates \cite{Cur98} conclude that vortex vibrations are
irrelevant in the relaxation of $^{63}$Cu but may play an important
role in that of $^{17}$O.

Recently, several groups have calculated $T_1$ in the superconducting state 
starting from a Fermi gas (FG) description of the cuprates, i.e.,  neglecting 
antiferromagnetic spin fluctuations \cite{Wor98,Vek98,Tak99}. We extended our 
calculations to the FG limit by using $\chi_{full} = \Pi/g^2$ 
and found two distinct differences between the relaxation rate for a FG 
and that in the presence of strong spin fluctuations. 
First, $1/T_1T$ for a FG always increases
with temperature as long as $A_{ab}$ and $B$  possess the same sign,
in contrast to our results in Eq.(\ref{T1Tf1}). 
Second, in the FG limit $1/T_1T$ increases monotonically with increasing 
local field in a vortex lattice,
i.e., it does not exhibit the local minimum we found in
Fig.~\ref{T1T_latt_T}. We propose that these differences in the
predicted relaxation rates enable NMR experiments to determine which
of the two limits applies to the superconducting state of the 
high-$T_c$ cuprates.

Finally, we found above that the behavior of the relaxation rate in a
vortex lattice reflects the presence of a supercurrent, magnetic
fluctuations and nodes in the superconducting gap. We are currently
studying \cite{Morr99a} whether similar effects also occur in
Sr$_2$RuO$_4$ for which strong indications of 
ferromagnetic fluctuations and a p-wave order parameter
exist \cite{Rice98}.   

In summary, we have demonstrated that NMR is a 
site-specific probe for the electronic structure in the mixed state of
the high-$T_c$ cuprates. We have shown that in a hexagonal vortex 
lattice, the Doppler-shifted electronic spectrum gives rise to a 
characteristic temperature and field dependence of the $^{63}$Cu
relaxation rate.  
We propose an NQR experiment in which the direction of a uniform
supercurrent with respect to the crystal lattice is varied, and
predict a unique angular dependence of $T_1$. Finally, we argue 
that our strong coupling results are qualitatively
different from those predicted for a FG. 

It is our pleasure to thank N. Curro, A. Leggett, C. Milling, D.
Pines,  J. Sauls, J. Schmalian, C. P. Slichter, J. Berlinsky and
C. Kallin for valuable
discussions and the Aspen Center for Physics for its hospitality in
the initial stages of this work. This work has been supported in part
by the Science and Technology Center for Superconductivity through
NSF-grant DMR91-20000 (D.K.M) and by the Natural Sciences and
Engineering Research Council of Canada (R.W.).

\end{document}